\documentclass[aps,pra,amsmath,amssymb,superscriptaddresss,12pt,showpacs]{revtex4}

\usepackage[sort&compress]{natbib}

\newcommand{\avg}[1]{\langle #1 \rangle}

\begin{document}

\title{A linear program for testing local realism}

\author{Matthew B.~Elliott}
\email{mabellio@gmail.com}

\begin{abstract}

We present a linear program that is capable of determining whether a set of correlations can be captured by a local realistic model. If the correlations can be described by such a model, the linear program outputs a joint probability distribution that produces the given correlations. If the correlations cannot be described under the assumption of local realism, the program outputs a Bell inequality violated by the correlations.

\end{abstract}


\maketitle

\section{Introduction}

Ever since the pivotal work by Bell \cite{bell:epr}, there has been a considerable effort to try and understand the quantum-mechanical violation of local realism (as a sample, see Refs.~\cite{acin:bell,braunstein:bell,cabello:bell,clauser:bell,gisin:bell,guhne:bell,laskowski:bell,masanes:bell,mermin:bell,werner:bell,zukowski:bell}). The typical approach in most work on the subject is to search for Bell inequalities~\cite{peres:bell}. Experiments to test the quantum-mechanical violation of local realism have then been designed with a particular Bell inequality in mind~\cite{aspect:lrexp,rowe:lrexp,weihs:lrexp}. In this paper we take a slightly different approach: given a set of correlations, we show how a relevant Bell inequality can be found. Thus, given an experimental setup, meaning the ability to create a particular quantum state and the ability to make a particular set of measurements, we illustrate how to derive a relevant Bell inequality which uses the given experimental setup to demonstrate a violation of local realism.

Peres has already presented a general approach to finding all the Bell inequalities given a set of measurement configurations~\cite{peres:bell}. In this paper, we reproduce this result via another computational method involving linear programming. Specifically, we present a linear program which can be used to generate all the Bell inequalities for a given set of measurement configurations. Linear programs constitute a class of widely applicable optimization problems for which many efficient algorithms exist~\cite{chvatal:book}. Moreover, there exists a rich theory of linear programming which we can exploit.

Our linear programming approach to problems involving local realism goes beyond finding all the Bell inequalities. Given an experimental setup, our linear program is capable of honing in on a single relevant Bell inequality. The linear program does not simply list all the Bell inequalities, of which there may be a large number, and present one that is violated. Instead, it \emph{directly} finds a Bell inequality that can be used to demonstrate a violation of local realism. Thus, rather than tailor an experiment to the available Bell inequalities, one can use our linear program to find a Bell inequality tailored to an experiment that can be carried out. This allows the possibility of using a more robust and experimentally feasible setup for which a relevant Bell inequality can then be found.

Refs.~\cite{basoalto:lp,zukowski:lp} also employ a linear program to find Bell inequalities. One major difference in our approach is in the event that a particular experimental setup is incapable of demonstrating a violation of local realism. In that case, no Bell inequality exists which is violated by the observed correlations, and our linear programming technique generates a local realistic description of all the observed correlations in the experiment. This is useful, for instance, in challenging claims of relevant Bell inequalities~\cite{collins:error}. We now move to demonstrate how all of this is accomplished.

Suppose, for simplicity, that we have two separated systems on which measurements of $A_j$, $1\leq j \leq m_A$, and $B_k$, $1\leq k \leq m_B$, are made. The generalization to more than two systems is straightforward. After the measurements are made, we can then compare the results and determine the experimentally observed correlations, $\avg{A_j B_k}$. If we allow the possibility of no measurement, meaning that $A_j = I$ or $B_k = I$, then these average values also include single system average values. If no measurement is made on either system, meaning $A_j = I$ and $B_k = I$, then we obtain the normalization condition. 

The existence of a local realistic description of the correlations depends on the existence of a joint probability distribution,
\begin{equation}
\label{eq:p}
P(A_1 \rightarrow a_1 , \ldots , A_{m_A} \rightarrow a_{m_A} ; B_1 \rightarrow b_1 , \ldots , B_{m_B} \rightarrow b_{m_B}) = P(\{A_j \rightarrow a_j\};\{B_j \rightarrow b_j\}),
\end{equation}
that gives the probability that a measurement of $A_j$ yields the result $a_j$ and a measurement of $B_k$ yields the result $b_k$. This probability distribution must be normalized,
\begin{equation}
\sum_{a_1,\ldots ,a_{m_A}} \sum_{b_1,\ldots ,b_{m_B}} P(\{A_j \rightarrow a_j\};\{B_k \rightarrow b_k\}) = 1,
\end{equation}
and non negative,
\begin{equation}
\forall a_j,b_k \hspace{1em} P(\{A_j \rightarrow a_j\};\{B_k \rightarrow b_k\}) \geq 0.
\end{equation}
It should also reproduce the observed correlations,
\begin{equation}
\label{eq:corrsum}
\avg{A_{j'} B_{k'}} = \sum_{a_1,\ldots ,a_{m_A}} \sum_{b_1,\ldots ,b_{m_B}} a_{j'} b_{k'} P(\{A_j \rightarrow a_j\} ; \{B_k \rightarrow b_k\}).
\end{equation}
The existence of this joint probability distribution is equivalent to the existence of a local realistic description of the observed correlations.

\section{Linear programming approach to local realism}

In order to see the applicability of linear programming to questions about local realism, we consider a column vector, $\vec{P}$, whose entries are the probabilities in Eq.~(\ref{eq:p}). In the common scenario of the measurements having two outcomes, this vector has dimension $2^{m_A+m_B}$. In terms of this vector notation for the joint measurement probabilities, the normalization and non-negativity constraints become, respectively,
\begin{equation}
\begin{array}{ccc}
\vec{1}^T \vec{P} = 1 & \mbox{and} & \vec{P} \geq \vec{0},
\end{array}
\end{equation}
where $\vec{1}$ and $\vec{0}$ are vectors of appropriate dimension with all entries equal to $1$ or $0$, and we use the linear programming convention that $\vec{A} \geq \vec{B}$ when each entry of $\vec{A}$ is greater than or equal to the corresponding entry of $\vec{B}$.

According to Eq.~(\ref{eq:corrsum}), any average value can be written as a linear combination of entries in $\vec{P}$. Let us write the average values that can be observed in some experiment, of which there will be $(m_A +1) (m_B +1)$ if we include no measurement, as entries in a vector, $\vec{C}$. Then this implies that a local realistic description of these correlations exists if and only if, for the appropriate matrix $M$, there exists a vector $\vec{P}$ with non-negative entries such that $M \vec{P} = \vec{C}$. In the case that measurements have two outcomes $\pm 1$, $M$ will be an $(m_A~+~1) (m_B~+1)~\times~2^{m_A+m_B}$ matrix whose entries are $\pm 1$. In general, $M$ is completely determined by the number of measurements and the number of outcomes, as well as their allowed values. Thus the problem is, given $M$ and $\vec{C}$, to solve
\begin{equation} \label{eq:solveprobdist}
\begin{array}{ccc}
M \vec{P} = \vec{C} & \mbox{s.t.} & \vec{P} \geq \vec{0}.
\end{array}
\end{equation}
Notice we do not include the normalization constraint separately because it is contained in the equation $M \vec{P} = \vec{C}$. In fact, we assume that the first row of $M$ has all $1$'s and that the first entry in $\vec{C}$ is $1$, as this takes care of normalization. Problems like these have been studied extensively and are called linear programs.

\subsection{Linear programming}

The most general form of a linear program is, given $\vec{L}$, $M$, and $\vec{C}$, to solve
\begin{equation} \label{eq:lp}
\begin{array}{cccccc}
\mbox{minimize} & \vec{L}^T \vec{P} & \mbox{s.t.} & M \vec{P} \geq \vec{C} & \mbox{and} & \vec{P} \geq \vec{0}.
\end{array}
\end{equation}
Just about any linear optimization problem involving linear constraints can be put into this form. We will see later how our problem can be put into this form, but first there is an important feature of linear programming needing discussion. For every linear program there is a related dual linear program,
\begin{equation} \label{eq:duallp}
\begin{array}{cccccc}
\mbox{maximize} & \vec{C}^T \vec{Q} & \mbox{s.t.} & M^T \vec{Q} \leq \vec{L} & \mbox{and} & \vec{Q} \geq \vec{0}.
\end{array}
\end{equation}
Notice that $\vec{P}$ appears nowhere in the dual program. This program seeks a completely different quantity $\vec{Q}$ that generally has a different dimension than $\vec{P}$. Yet there is an important relationship between these problems. The strong duality theorem for linear programming states that the minimum value of $\vec{L}^T \vec{P}$ in Eq.~(\ref{eq:lp}) is exactly equal to the maximum value of $\vec{C}^T \vec{Q}$ in Eq.~(\ref{eq:duallp})~\cite{chvatal:book}. We will see that duality in linear programming reveals a duality between probability distributions and Bell inequalities.

We now make a few comments regarding the theory of linear programs. First, the complexity of the optimization problems in Eqs.~(\ref{eq:lp}) and~(\ref{eq:duallp}) is polynomial in the dimensions of $\vec{L}$, $M$, and $\vec{C}$. Also, the optimal solutions found are not unique, as there may be several vectors achieving the optimal value for a given linear program.

\subsection{A linear program for probability distributions}

Returning to Eq.~(\ref{eq:solveprobdist}), we try to rewrite our problem as a linear program,
\begin{equation}
\begin{array}{cccccc}
\mbox{minimize} & \vec{1}^{\phantom{.}T} (M \vec{P} - \vec{C}) & \mbox{s.t.} & M \vec{P} \geq \vec{C} & \mbox{and} & \vec{P} \geq \vec{0}.
\end{array}
\end{equation}
A solution to Eq.~(\ref{eq:solveprobdist}) exists if and only if this minimum is $0$. The reason is that $M \vec{P} \geq \vec{C}$ assures that every entry of $M \vec{P} - \vec{C}$ is non negative. Thus the sum of all the entries in $M \vec{P} - \vec{C}$ is $0$ if and only if each entry of $M \vec{P} - \vec{C}$ is zero, i.e $M \vec{P} = \vec{C}$. Now we easily turn this into a linear program,
\begin{equation} \label{eq:lpprobdist}
\begin{array}{cccccc}
\mbox{minimize} & (M^T \vec{1}\phantom{.})^T \vec{P} & \mbox{s.t.} & M \vec{P} \geq \vec{C} & \mbox{and} & \vec{P} \geq \vec{0}.
\end{array}
\end{equation}
If the minimum value is $\vec{1}^{\phantom{.}T} \vec{C}$, then there exists a probability distribution, $\vec{P}$, producing the given measurement correlations, $\vec{C}$. Otherwise, no such probability distribution exists.

This linear program gives us a method for determining whether a set of correlations violates local realism. In addition, if the correlations can be reproduced in a local and realistic way, then the output, $\vec{P}$, of the program constitutes such a description.

\subsection{A linear program for Bell inequalities}

There is a deep connection between probability distributions and Bell inequalities that arises from duality in linear programming. The dual program of Eq.~(\ref{eq:lpprobdist}), which produces Bell inequalities, is
\begin{equation} \label{eq:duallpbi}
\begin{array}{cccccc}
\mbox{maximize} & \vec{C}^T \vec{Q} & \mbox{s.t.} & M^T \vec{Q} \leq M^T \vec{1} & \mbox{and} & \vec{Q} \geq \vec{0}.
\end{array}
\end{equation}
From the strong duality theorem we know that there exists a local realistic description of the correlations in $\vec{C}$ if and only if this maximum value is $\vec{1}^{\phantom{.}T} \vec{C}$. Since $\vec{Q} = \vec{1}$ satisfies the constraints and achieves this value, we know that generally the maximum value must exceed $\vec{1}^{\phantom{.}T} \vec{C}$.

From this we deduce that there exists a local realistic description for a set of correlations if and only if the linear program,
\begin{equation} \label{eq:bi}
\begin{array}{cccccc}
\mbox{maximize} & \vec{C}^T (\vec{Q}-\vec{1}) & \mbox{s.t.} & M^T (\vec{Q}-\vec{1}) \leq \vec{0} & \mbox{and} & (\vec{Q}-\vec{1}) \geq -\vec{1}
\end{array}
\end{equation}
outputs $0$. This gives us the set of Bell inequalities $\vec{C}^T (\vec{Q}-\vec{1}) \leq 0$ for all vectors $\vec{Q}-\vec{1}$ satisfying the constraints. This is a complete set of Bell inequalities in that they are both necessary and sufficient for a local realistic description to exist. They all have the form of a linear combination of correlations being less than or equal to $0$.

It may appear that Eq.~(\ref{eq:bi}) constitutes an infinite number of Bell inequalities. However, notice that the vectors satisfying the constraints form a convex set with a finite number of extreme points. These extreme points, by themselves, then constitute a finite set of Bell inequalities which are necessary and sufficient for the given correlations to be described in a local realistic framework. Thus we do not have an infinite number of Bell inequalities, but only a number equal to the number of extremal vectors in the constraint set. Moreover, this complete set of Bell inequalities is independent of $\vec{C}$, the correlations, and can be determined given only the matrix $M$, which is in turn only a function of the number of measurements as well as the possible measurement outcomes.

If we run the dual program in Eq.~(\ref{eq:duallpbi}) and find a vector $\vec{Q}$ that satisfies the constraints, but such that $\vec{C}^T \vec{Q} > \vec{1}^T \vec{C}$, then we have a Bell inequality, $\vec{C}^T (\vec{Q}-\vec{1}) \leq 0$, that is violated by the given correlations in $\vec{C}$. At this point we have completed what we set out to do: given a set of correlations, we have a linear program that either finds a local realistic description of them or produces a Bell inequality that they violate.

As a final exercise, we show how these Bell inequalities can be simplified to yield a standard form for Bell inequalities. The first thing is to recognize that the constraint $(\vec{Q}-\vec{1}) \geq -\vec{1}$ is not needed. For, suppose that we have a vector $\vec{Q}-\vec{1}$ satisfying $M^T (\vec{Q}-\vec{1}) \leq \vec{0}$, but such that $(\vec{Q}-\vec{1}) \geq -q \vec{1}$ for some $q > 1$. Then the vector $(\vec{Q}-\vec{1})/q$ satisfies all constraints, and thus we get the Bell inequality $\vec{C}^T (\vec{Q}-\vec{1})/q \leq 0 \Leftrightarrow \vec{C}^T (\vec{Q}-\vec{1}) \leq 0$. So in fact we get a Bell inequality for all vectors $\vec{Q}-\vec{1}$ satisfying $M^T (\vec{Q}-\vec{1}) \leq \vec{0}$, regardless of whether they satisfy the second constraint.

A second simplification can also be used to deal with the constraint $M^T (\vec{Q}-\vec{1}) \leq \vec{0}$. Since we assume the first row of $M$ is all $1$'s, corresponding to the normalization of $\vec{P}$, we can break up $\vec{Q}-\vec{1}$ as $\vec{Q} - \vec{1} = \left( \begin{array}{cc} -q_0 & \vec{q} \end{array} \right)^T$, where $\vec{q}$ has one less dimension than $\vec{Q}$. If we also break up $\vec{C} = \left( \begin{array}{cc} 1 & \vec{c} \end{array} \right)^T$, then the Bell inequalities are $-q_0+\vec{c}^{\phantom{.}T} \vec{q} \leq 0$ for all $-q_0 \vec{1} + \tilde{M}^T \vec{q} \leq \vec{0}$, where $\tilde{M}$ is $M$ without the first row. To have a complete set of Bell inequalities it suffices to pick the smallest $q_0$ and so we have $q_0 = (\tilde{M}^T \vec{q})_{\mbox{\scriptsize{max}}}$, where the maximum denotes the largest entry of the vector, giving us all the Bell inequalities,
\begin{equation}
\begin{array}{ccc}
\vec{c}^{\phantom{.}T} \vec{q} \leq (\tilde{M}^T \vec{q})_{\mbox{\scriptsize{max}}} & \mbox{for all} & \vec{q}.
\end{array}
\end{equation}
This is often how Bell inequalities are stated; all the Bell inequalities are obtained by considering an arbitrary linear combination of correlations and then saying that this must be less than the maximum possible value of those correlations under extremal local realistic predictions.

\section{Conclusion}

What we have shown here is a computational method that determines whether a set of correlations can be described within a local realistic framework. If the set is so describable, then the linear program gives a probability distribution resulting in those correlations. If the set is not describable by a local realistic model, then the linear program is capable of generating a Bell inequality that the correlations violate.

As a practical matter, this program can be used to generate a relevant Bell inequality for a particular Bell test experiment. In other words, it takes some of the difficulty away from designing an experiment to test the quantum violation of local realism and puts that difficulty into a program that yields a relevant Bell inequality to study. This is particularly useful in the multipartite scenario where relevant Bell inequalities are less plentiful. In terms of theory, linear programming gives a way to generate all the Bell inequalities and through the principle of duality in linear programming we see that probability distributions and Bell inequalities are also dual to each other.

We note that the linear program presented here scales exponentially in the number of systems and the number of measurements made on the systems. This, of course, is not surprising because this problem is known to be NP hard~\cite{peres:bell}. For a reasonable number of systems and measurements, however, such as could be managed in an experiment, this program should still be of use.

As a final observation, since the source of the correlations, $\vec{C}$, is irrelevant, our linear program is also applicable to correlations that go beyond measurements on quantum states. In fact, one can consider the framework of generalized probabilistic theories~\cite{barrett:theories}, which encompasses not only quantum correlations but also the stronger correlations of PR boxes~\cite{popescu:prbox}, and determine whether a set of given correlations can be reproduced in a local realistic framework.

\acknowledgments

Thanks to H.~N.~Barnum, I.~H.~Deutsch, and A.~J.~Landahl for their encouragement and careful readings of a previous version of this document. Special thanks to C.~M.~Caves for particularly helpful comments and suggestions. The work presented here was supported by National Science Foundation Grant No.~PHY-0653596.

\end{document}